\documentclass[11pt]{article}

\usepackage[dvips]{epsfig}
\usepackage{amssymb}
\usepackage{multirow}
\usepackage{amsbsy}
\usepackage{amsmath}
\usepackage{pstricks}

\usepackage{authblk}

\newcommand{\mathd}{\mathrm{d}}
\newcommand{\mathe}{\mathrm{e}}
\newcommand{\nobracket}{}
\newcommand{\tmop}[1]{\ensuremath{\operatorname{#1}}}
\newcommand{\eq}{Eq.~}
\newcommand{\eqs}{Eqs.~}
\newcommand{\fig}{Fig.~}

\newcommand{\cf} {cf.~}
\newcommand{\ug} {\!=\!}
\newcommand{\piu} {\!+\!}
\newcommand{\meno} {\!-\!}
\newcommand{\ie} {i.e.~}
\newcommand{\eg} {e.g.~}
\newcommand{\rref} {Ref.~}
\newcommand{\rrefs} {Refs.~}

\newcommand{\U}{\mathcal{U}_\tau}
\newcommand{\J}{\mathcal{J}}
\newcommand{\R}{\mathcal{R}}

\title{Quantum non-Markovian piecewise dynamics from collision models}

\author[1,2]{Salvatore Lorenzo}
\author[3,4,5]{Francesco Ciccarello}
\author[3,4]{G. Massimo Palma}
\author[1,2]{Bassano Vacchini}

\affil[1]{Quantum Technology Lab, Dipartimento di Fisica, Universit\`a  degli Studi di Milano, I-20133 Milano, Italy} 
\affil[2]{INFN, Sezione di Milano, I-20133 Milano, Italy}
\affil[3]{Dipartimento di Fisica e Chimica, Universit\`a  degli Studi di Palermo, via Archirafi 36, I-90123 Palermo, Italy}
\affil[4]{NEST, Istituto Nanoscienze-CNR}
\affil[5]{Department of Physics, Duke University, P.O. Box 90305, Durham, North Carolina 27708-0305, USA}

\begin{document}

\title{Quantum non-Markovian piecewise dynamics from collision models}

\maketitle

\begin{abstract}
Recently, a large class of quantum non-Markovian piecewise
  dynamics for an open quantum system obeying closed evolution
  equations has been introduced [B. Vacchini, Phys. Rev. Lett. {\bf
    117}, 230401 (2016)]. These dynamics have been defined in terms of
  a waiting-time distribution between quantum jumps, along with quantum
  maps describing the effect of jumps and the system's evolution
  between them. Here, we present a quantum collision model with
  memory, whose reduced dynamics in the continuous-time limit
  reproduces the above class of non-Markovian piecewise dynamics, thus
  providing an explicit microscopic realization.
\end{abstract}


\date{}

\section{Introduction} \label{intro}

Prompted by the growing impact of quantum technologies, the study of
non-Markovian (NM) quantum dynamics is currently a topical field
\cite{review-breuer,review-plenio,review-vacchini,review-vega}. Besides
the goal of defining, witnessing and even quantifying on a rigorous
basis the degree of quantum ``non-Markovianity" of an open system dynamics,
efforts are under way to advance the longstanding quest for the non-Markovian
counterpart of the celebrated Gorini-Kossakowski-Lindblad-Sudarshan master equation (ME)
\cite{Gorini1976a,Lindblad1976a}. As a pivotal requisite, which may be
easily violated \cite{barnettPRA2001,maniscalcoPRA2006}, a
well-defined NM ME must entail a completely positive and
trace-preserving (CPT) dynamics {for an arbitrary initial state and for
suitably large classes of operators and parameters appearing in its
expression.} While the set of known NM dynamics described by
well-behaved MEs (in the above sense) is still relatively small,
remarkable progress was made in the last few years.

A relatively new approach to quantum NM dynamics is based on {quantum collision models (CMs)}
\cite{rybarJPB2012,giovannettiPRL2012,giovannettiJPB2012,ciccarelloPRA2013,santosPRA2014,grimsmoPRL2015,lorenzoPRA2016,luomaPRA2016,lorenzoARX2017}. A CM is a simple microscopic framework for describing the open dynamics of a system $S$ in contact with a bath, where the latter is assumed to consist of a large number of elementary subsystems, the ``ancillas". The open dynamics of $S$ results  from its successive pairwise {\it collisions} with the bath ancillas, each collision being typically described by a bipartite unitary on $S$ and the involved ancilla. 

In the continuous-time limit, a CM leads to a Lindblad ME with no need to resort to the Born-Markov approximation \cite{buzekOSID2005}. Such appealing property prompted NM generalizations of the simplest memoryless CM, whose continuous-time-limit dynamics is ensured by construction to be CPT. A significant instance is the CM in \rrefs\cite{ciccarelloPRA2013,ciccarelloPS2013}, recently extended in \rref\cite{lorenzoPRA2016}, which produced a new NM memory-kernel ME. The peculiar structure of this memory-kernel ME and corresponding dynamical map inspired further investigations \cite{vacchiniPRA2013,vacchiniIJQI2014,dariusPRA2016,vacchiniPRL2016,Chruscinski2017a} from different {\it viewpoints}, which allowed to further enlarge the class of known NM dynamics governed by well-defined MEs. 

One of these viewpoints builds on the well-known quantum-jumps picture
of the Lindblad ME \cite{petruccione-book,Holevo2001,Barchielli2009,Barchielli1994a} to devise a far larger,
NM class of piecewise dynamics characterized by a waiting time
distribution, a CPT map describing the effect of jumps and a
collection of CPT maps accounting for the evolution between jumps
\rref\cite{vacchiniPRA2013,vacchiniIJQI2014,vacchiniPRL2016}. This
class of piecewise dynamics obeys a memory-kernel ME
\cite{vacchiniPRL2016}. Given that this general ME encompasses the
reduced ME of the CM in \rref\cite{lorenzoPRA2016} only as a special
case, it is natural to wonder whether a generalized CM can be
constructed giving rise to the piecewise-dynamics ME with no
restrictions. In this work, we prove that such a CM indeed exists and
show that it can be defined as a non-trivial generalization of
\rref\cite{lorenzoPRA2016} where collisions occur in the form of
probabilistic SWAP operations. Among its major distinctive features
are the {\it doubling} of each ancilla into a pair of subancillas,
which allows to introduce the jump map that was fully absent in
\rref\cite{lorenzoPRA2016}, and the introduction of {{\it
  time-step-dependent}} swap probabilities, which allows to reproduce
waiting time distributions of arbitrary shape unlike
\rref\cite{lorenzoPRA2016} that was restricted to exponential
ones. This extension is of particular importance to comply with
possible experimental implementations as well as encompass all the different
features of the interaction dynamics that might give rise to
non-Markovianity.

This paper is organized as follows. In Section \ref{review1}, we review the class of NM quantum dynamics introduced in \rrefs\cite{vacchiniPRA2013,vacchiniIJQI2014,vacchiniPRL2016}. As anticipated, the main purpose of this work is demonstrating the existence of a quantum CM with memory, which in the continuous-time limit reproduces the above class of NM dynamics. Since this CM is an extension of the one in \rref\cite{lorenzoPRA2016}, the latter is reviewed in Section \ref{review2} and a brief introduction to quantum CMs is provided. These introductory sections, in particular, allow us to introduce most of the notation and formalism that we use later in Section \ref{CM-model}, where the main results of this work are presented. Owing to its central importance, Section \ref{CM-model} is structured in a number of subsections so as to better highlight the different essential aspects of the proposed CM: the initial state, the way system-ancilla collisions are modelled, the discrete dynamics, its continuous-time limit and, at last, the reduced dynamics of the open system. Our conclusions along with some comments and outlook are given in Section \ref{conc} Some technical proofs are presented in Appendix A.

\section{Review of non-Markovian piecewise quantum dynamics} \label{review1}

The prototypical Markovian dynamics of an open quantum system $S$ is
described by the Gorini-Kossakowski-Lindblad-Sudarshan ME \cite{Gorini1976a,Lindblad1976a}, which reads
\begin{equation}
\dot{\rho}=-i[\hat H,\rho]+\sum_{k}\gamma_k \left(\hat L_{k}\rho \hat L^\dagger_{k}-\frac{1}{2}\lbrace \hat L^\dagger_{k}\hat L_{k},\rho\rbrace\right)\,,
\label{LME} \end{equation}
where $\rho(t)$ is the $S$ density operator, $\{...,...\}$ stands for the anticommutator, $\hat H$ is a Hermitian operator, $\{\gamma_k\}$ are positive rates, and where $\{\hat L_k\}$ are jump operators.
By introducing the maps
\begin{eqnarray}
\R_t[\rho]=e^{\hat Rt}\rho\, e^{\hat R^\dagger t}\,,\,\,\,\,\,\, \J[\rho]=\sum_{k}\gamma_k\,\hat L_{k}\rho \hat L^\dagger_{k}\,,\label{R-J}
\end{eqnarray}
where we defined the non-Hermitian operator $\hat R=-i \hat
H-\frac{1}{2}\sum_k\gamma_k\hat L^\dagger_{k}\hat L_{k}$, the solution
of the Lindblad ME (\ref{LME}) can be written as the Dyson series \cite{Holevo2001}
\begin{eqnarray}
  \rho_t = {\mathcal{R}}_{t} [ \rho_0]
 + \sum_{j=1}^{\infty} \!\int_{0}^{t} \hspace{-0.17em} \hspace{-0.17em}
  \tmop{d}\!t_{j} \ldots \hspace{0.27em} \!\int_{0}^{t_{2}} \hspace{-0.17em}
  \hspace{-0.17em} \tmop{d}\!t_{1} \hspace{0.27em} \ldots  \mathcal{R}_{t-t_{j}} \mathcal{J} \ldots
  \mathcal{J} \mathcal{R}_{t_{2} -t_{1}} \mathcal{J}
{\mathcal{R}}_{t_{1}} [ \rho_0 ] \nonumber\\\,\,\,\,\,\,\,\label{dyson}
\end{eqnarray}
with $0\le t_1\le t_2\le...\le t$.
\eq(\ref{dyson}) shows that the time evolution of $S$ can be viewed as
an underlying dynamics described by the evolution map $\mathcal R_t$
interrupted by jumps each transforming the system state according to
the jump map $\mathcal J$. Index $j$ in \eq(\ref{dyson}) indeed
represents the number of jumps occurred up to time $t$ at instants
$\{t_1,t_2,...,t_j\}$ such that $0\le t_1\le t_2\le...\le t_j\le
t$. Note that the maps (\ref{R-J}) are {\it not} trace-preserving.

Both the Lindblad ME \eq(\ref{LME}) and the representation \eq(\ref{dyson}) for
its exact solution have been taken as a starting point for possible
generalizations leading to well-defined dynamics to be described by
means of memory kernel MEs, which can describe memory effects in
the time evolution. Starting from the seminal work in \rref \cite{Budini2004a},
different approaches have been devised along this line
\cite{Budini2005a,Breuer2008a,Kossakowski2009a,Budini2013a,Budini2013b}.
One of us recently extended these results investigating a NM
generalization of \eq(\ref{dyson})
\cite{vacchiniPRA2013,vacchiniIJQI2014,vacchiniPRL2016}, which in its
most general form can be
expressed as \cite{vacchiniPRL2016}
\begin{eqnarray}
  \rho_t&{=}& g ( t ) \bar {\mathcal{E}}_{t} [ \rho_0  ] + \sum_{j=1}^{\infty} \int_{0}^{t} \hspace{-0.17em} \hspace{-0.17em}
  \tmop{d}\!t_{j} \ldots \hspace{0.27em} \int_{0}^{t_{2}} \hspace{-0.17em}
  \hspace{-0.17em} \tmop{d}\!t_{1} \hspace{0.27em}   \nonumber\\&&
  f ( t{-}t_{j} ) \ldots f (
  t_{2}{-}t_{1} ) g ( t_{1} ) \mathcal{E}_{t-t_{j}} \mathcal{Z} \ldots
  \mathcal{Z} \mathcal{E}_{t_{2} -t_{1}} \mathcal{Z}
  \bar {\mathcal{E}}_{t_{1}} [ \rho_0  ]. \nonumber\\
  \label{dysonNM}
\end{eqnarray}
Compared to \eq(\ref{dyson}), the jump map $\cal J$ [see
\eq(\ref{R-J})] is turned into the CPT map $\cal Z$, while ${\mathcal
  R}_t$ is replaced by the CPT evolution map $\bar {\mathcal E}_t$
{\it before} any jump has occurred and by the CPT evolution map
$\mathcal E_t$ {\it after} the first jump (if any) has taken
place. Maps $\cal Z$, $\bar {\mathcal E}_t$ and $\mathcal E_t$ are
fully unspecified, but for the requirement of being CPT. Importantly, while in \eq(\ref{dyson}) the statistical weight of each possible trajectory is determined by the non-trace-preserving maps $\mathcal R_t$, $\mathcal J$ and the initial state \cite{vacchiniIJQI2014}, 
in \eq(\ref{dysonNM}) these statistical weights are assigned
independently of the maps $\bar {\mathcal E}_t$, ${\mathcal E}_t$, $\cal Z$ and the initial state.
Indeed, the functions $f(t)$ and $g(t)$, appearing in
\eq(\ref{dysonNM}), stand for an arbitrarily chosen {\it waiting time distribution}, namely the probability 
density for the distribution in time of the jumps, and its associated
{\it survival probability} $g(t){=}1{-}\int_0^t {\rm d}t' f(t')$, that
is the probability that no jump has taken place up to time $t$. 
The waiting time distribution and the associated survival probability can always be expressed in the form
\begin{eqnarray}
  g ( t ) {=}  \exp \left[ - \int^{t}_{0} \mathd s \phi ( s ) \right]
 \,,\,\,\,\,
  f ( t )  {=}  \phi ( t )\, \exp \left[ - \int^{t}_{0} \mathd s \phi ( s )
  \right]\,, 
\label{f-and-g}
\end{eqnarray}
where the positive function
\begin{eqnarray}
  \phi (t) & = & \frac{f ( t )}{g ( t )} 
\label{hazard}\end{eqnarray}
is known as {\it hazard rate function} or simply {\it hazard function}
\cite{Ross2007}. 
The meaning of this is that $\phi ( t ) \tmop{dt}$ provides
the probability for a jump to take place in the time interval $( t,t+
\tmop{dt} ]$, given that no jump has taken place up to time $t$.
Accordingly, the time-dependent coefficient $ f ( t_{} \!-\!t_{j} )
\!\ldots\! f (t_{2} \!-\!t_{1} ) g ( t_{1} )$ in \eq(\ref{dysonNM})
gives the probability density that $j$ jumps take place at times
$\{t_1,t_2,...,t_j\}$, while the pre-factor of the first term on the
rhs is the probability that no jumps occurred up to time $t$ (this
pre-factor indeed multiplies the jump-free evolution map $\bar
{\mathcal E}_t$). The jumps are thus distributed in time
  according to a renewal process, which in particular entails that
  after each jump the process starts anew. Note that, in fact by construction, the dynamical map defined by \eq(\ref{dysonNM}) is ensured to be CPT. Importantly, it can be shown \cite{vacchiniPRL2016} that it obeys the memory-kernel ME
\begin{eqnarray}
\dot{\rho} =\int_{0}^{t} {\rm d}t'\,\mathcal{W}({t-t'})[\rho(t')]+\mathcal{I}(t)[\rho_0 ]\,,\label{ME-piecewise}
\end{eqnarray}
where
\begin{eqnarray}
\mathcal{W}(t)=\frac{\rm d}{{\rm d}t}\left[f(t)\mathcal{E}_t\right]\mathcal Z+\delta(t)f(0)\mathcal{E}_0\mathcal Z\,,\,\,\,\,\,\,
\mathcal{I}(t)=\frac{\rm d}{{\rm d}t}\left[g(t)\bar {\mathcal E}_t\right]\,.
\end{eqnarray}
The corresponding open dynamics, at variance with the Lindbladian case [see \eqs(\ref{LME}) and (\ref{dyson})], is in general NM \cite{vacchiniPRA2013}.

\section{Collision models with memory} \label{review2}
\setcounter{equation}{0}

A quantum CM \cite{rauPR1963,scaraniPRL2002,zimanPRA2002} is a simple microscopic model for describing the open dynamics of a system $S$ in contact with a bath $B$. In its prototypical version, a CM assumes that $B$ comprises a huge number of elementary, identical and non-interacting ancillas all initialized in the same state $\eta$. The $S$-$B$ interaction process occurs via successive pairwise ``collisions" between $S$ and the ancillas, each of these collisions being described by a bipartite unitary operation $\hat U_{n}$. 
By hypothesis, $S$ can collide with each ancilla only once. After $n$ collisions, the state of $S$ is given by $\rho_n{=}\Phi^n [\rho_0]$, where the CPT map $\Phi$ is defined as 
$\Phi[\rho]{=}{\rm Tr}_n \{\hat U_{n}(\rho {\otimes}\eta_n)\hat U_{n}^\dagger\}$. 
Note that, despite the apparent dependance on $n$  (see \eg the
partial trace over the $n$th ancilla), the map $\Phi$ does not depend on $n$ since the bath initial state and system-ancilla interaction Hamiltonian are fully homogeneous. It can be shown \cite{buzekOSID2005} that in the continuous-time limit the dynamics of such a simple CM is described by a Lindblad ME of the form (\ref{LME}), a result which can be expected based on the discrete semigroup property enjoyed by the collision map, $\Phi^{n+m}=\Phi^n\Phi^m$. The open dynamics of $S$ corresponding to such a paradigmatic CM is thereby fully Markovian. 

There are several ways to endow the basic CM just described with memory so as to give rise to a NM dynamics \cite{rybarJPB2012,ciccarelloPRA2013,santosPRA2014,grimsmoPRL2015,luomaPRA2016,lorenzoARX2017}. The one of concern to us, given the goals of the present paper, is the CM with memory of \rref\cite{lorenzoPRA2016}, which can be regarded as a generalization of a model first put forward in \rrefs\cite{ciccarelloPRA2013,ciccarelloPS2013}. The general structure of the CM in \rref\cite{lorenzoPRA2016} is in many respects analogous to the basic memoryless CM described in the previous paragraph except that the system undergoing collisions with the bath ancillas is now bipartite, comprising the very open system under study $S$ {\it plus} an auxiliary system $M$, the ``memory", whose Hilbert space dimension is the same as each ancilla's one. A sketch of the CM is given in \fig1(a). Systems $S$ and $M$ interact all the time according to the pairwise unitary evolution map
\begin{equation}
\mathcal U_\tau[\sigma]{=} e^{-i \hat H_{S\!M} \tau}\sigma e^{i \hat H_{S\!M} \tau}\,,\label{mapU}
\end{equation}
where $\hat H_{S\!M}$ is the joint $S$-$M$ Hamiltonian. Here and throughout this paper, $\sigma$ stands for a joint state of the $S$-$M$ system {\it and} all the bath ancillas.
\begin{figure*}[htbp]
\begin{center}
\includegraphics[width=0.9\textwidth]{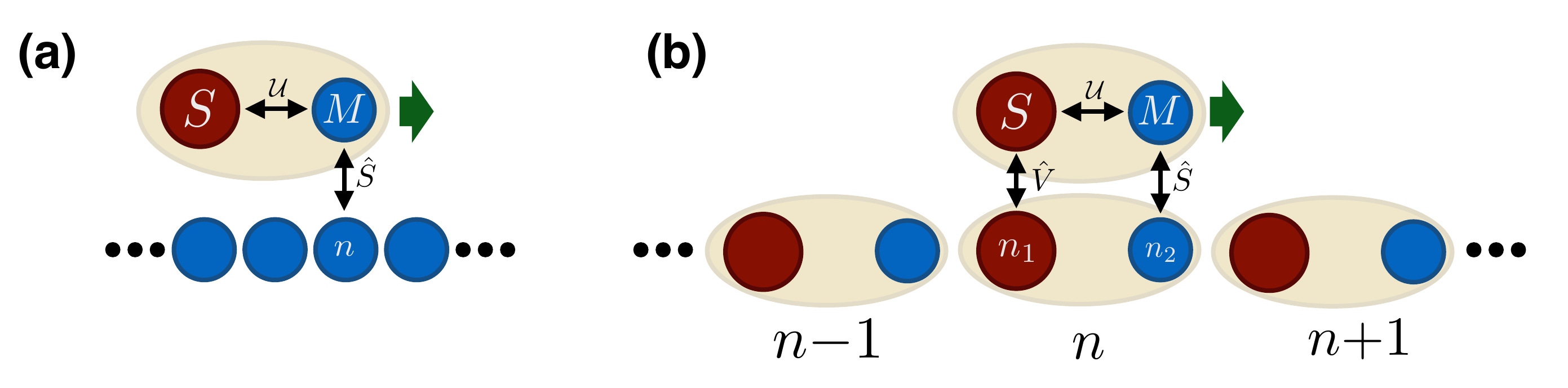}
\caption{(a) 
Sketch of the considered CMs: the system
  undergoing collisions with the bath ancillas is bipartite,
  comprising the very open system $S$ and the memory $M$, with the
  latter having the same dimension as each bath ancilla. Only $M$ is
  directly involved in collisions with ancillas. Each collision swaps
  the states of $M$ and the $n$th ancilla in a probabilistic way by
  means of the transformation $\hat S$ acting on $M$ and $n$. (b) Sketch
  of the generalized CM with memory of Section \ref{CM-model}:
  ancillas are now bipartite, the $n$th of which comprising a
  subancilla $n_1$ ($n_2$) having the same dimension as $S$ ($M$). Now
  both $S$ and $M$ are directly involved in collisions with
  ancillas. At each collision with some probability the states of
 $M$ and $n_2$ are swapped, and at the same time the bipartite unitary
$\hat V$ is applied on $S$ and $n_1$.}
\label{default}
\end{center}
\end{figure*}
By hypothesis, only $M$ is in direct contact with the bath [see \fig1(a)]. This interaction takes place through successive collisions, each being described by the pairwise {\it non-unitary} quantum map
\begin{equation}
\mathcal {S}_n[\sigma]{=}p\,  \sigma{+} (1{-}p)\hat{S}_{Mn}\sigma\hat{S}_{Mn}\,,\label{PS}
\end{equation}
where $\hat{S}_{Mn}$ is the swap unitary operator exchanging the states of $M$ and the $n$th ancilla. The CPT map (\ref{PS}), which depends parametrically on the probability $p$, can be interpreted as a probabilistic partial SWAP gate: the memory and ancilla states are either swapped or left unchanged with probability $p$.

The initial state of the overall system ($S$, $M$ and the bath ancillas) is assumed to be
\begin{equation}
\sigma_0=\left(\rho_0\otimes\bar\eta_M\right)\otimes\!\eta_1\!\otimes\!\eta_2\otimes\cdot\!\cdot\!\cdot\,\,\,\label{sigma0-bi}
\end{equation}
where in particular $\rho_0$ ($\bar\eta_M$) is the initial state of
$S$ ($M$). Throughout this paper, tensor product symbols will be
omitted whenever possible to avoid using too cumbersome notation.

By calling $\sigma_n$ the overall state at step $n$, the dynamics proceeds according to 
\begin{equation}
\sigma_n=\mathcal {S}_n\;\mathcal U_\tau\;\mathcal {S}_{n{-}1}\;\mathcal U_\tau\ldots\mathcal {S}_{{}_2}\;\mathcal U_\tau\;\mathcal{S}_{{}_1}\;\mathcal U_\tau[\sigma_0]\,,\label{sigman}
\end{equation}
namely an $S$-$M$ unitary dynamics goes on all the time, being interrupted at each {\it fixed} time step $\tau$ by a collision between $M$ and a  ``fresh" ancilla (\ie one still in the initial state $\eta$) described by the non-unitary map (\ref{PS}). Equivalently, one can view each $\mathcal U_\tau$ itself as embodying the effect of a unitary collision that is however internal to the joint $S$-$M$ system in such a way that the overall CM dynamics results from subsequent $M$-ancilla collisions interspersed with internal ones that involve $S$ and $M$ only \cite{lorenzoPRA2016}.

Like for any CM, the dynamics just defined is discrete. One can however define a {\it continuous-time limit} by assuming that the duration of each time step $\tau$ becomes very small while the step number $n$ gets very large in such a way that $n\tau\!\rightarrow t$, where $t$ is a continuous time variable. The assumption of a very large number of steps demands an additional prescription for the continuous-time limit of the probability $p$ entering \eq(\ref{PS}) since \eq(\ref{sigman}) clearly features $p$'s powers $\{p^k\}$ for all positive integers $k{\leq} n$. This task is carried out by first defining a rate $\Gamma$ that allows to express $p$ as
\begin{equation}
p= e^{-\Gamma \tau}\label{p-Gamma}\,
\end{equation}
(which is always possible) and assuming next that $\Gamma\tau\!\ll\!1$ in such a way that $p\simeq 1$. This ensures that $p^k$, for any $k$ smaller than $n$ and yet large enough so that $k \tau\rightarrow t'<t$ is finite, be not washed out in the continuous-time limit. Indeed, this yields
\begin{equation}
p^{k}\ug ({p^{\frac{1}{\tau}}})^{k\tau}\rightarrow e^{-\Gamma t'}\,.
\end{equation}
By finally noting that, consistently with the hypothesis $\Gamma\tau\ll 1$, $1{-} p\ug1{-} e^{-\Gamma \tau}{\simeq}\Gamma \tau$ [\cf \eq(\ref{PS})] and that 
since the CM is well defined for any choice of $\eta$, $\bar\eta$ and $\mathcal{U}_t$ 
it is possible to describe the reduced evolution of $S$ by the following CPT map
\begin{equation}
\rho_t{=} e^{-\Gamma t}\bar {\mathcal{E}}_{t} [ \rho_0  ] + \sum_{j=1}^{\infty} \Gamma^j e^{-\Gamma t}\int_{0}^{t} \tmop{d}\!t_{j} \ldots \int_{0}^{t_{2}} \tmop{d}\!t_{1}
\mathcal{E}_{t-t_{j}}\ldots\mathcal{E}_{t_{2} -t_{1}} \bar {\mathcal{E}}_{t_{1}} [\rho_0]. 
\label{partial-swap-map}\end{equation}
which is a special case of (\ref{dysonNM}) for 
\begin{eqnarray}
&\bar{\mathcal{E}}_{t}\rho={\rm Tr}_{M}\left\{\mathcal{U}_t[\rho \,\bar\eta_M]\right\},\,\,\,{\mathcal{E}}_{t}\rho={\rm Tr}_{M}\left\{\mathcal{U}_t[\rho \,\eta_M]\right\},\,\,\,\mathcal Z={\mathbb I}&,\,\,\,\nonumber \\
&f(t)=\Gamma e^{-\Gamma t},\,\,\,\, g(t)=e^{-\Gamma t}, \,\,\,\, \phi(t)=\Gamma.&\,\,\,\,\,\,\,\,
\label{special-case}
\end{eqnarray}
In fact it can be shown \cite{lorenzoPRA2016} that  the map (\ref{partial-swap-map}) obeys a memory-kernel ME of the form (\ref{ME-piecewise}).

Yet, the CM in fact lacks the jump map $\mathcal Z$ and is, in addition, apparently constrained to a purely exponential waiting time distribution $f(t){=}\Gamma e^{-\Gamma t}$ [the corresponding hazard function $\phi(t)$ being thus constant].

In the next section, we show how to construct a CM with memory whose continuous-time limit yields ME (\ref{ME-piecewise}) in the most general case, including in particular an arbitrary jump map $\mathcal Z$ and an arbitrary waiting time distribution $f(t)$.

\section{A generalized collision model with memory}\label{CM-model}

The CM to be defined here is a non-trivial generalization of the CM of \rref\cite{lorenzoPRA2016} reviewed in the last section. Just like in \rref\cite{lorenzoPRA2016}, the system undergoing collisions with the bath ancillas comprises $S$ and a memory $M$ that are subject to a coherent mutual coupling giving rise to the unitary evolution map (\ref{mapU}). At variance with \rref\cite{lorenzoPRA2016}, however, now each bath ancilla is {\it bipartite} as well, consisting of a pair of ``subancillas": one subancilla has the same Hilbert space dimension as $S$, while the other subancilla has the same dimension as $M$. A sketch of this generalized CM with memory is displayed in \fig1(b).

\subsection{Initial state}

The initial joint state reads
\begin{equation}
\sigma_0=\left(\rho_0\otimes\bar\eta_M\right)\otimes\left(\xi_1\otimes\eta_1\right)\otimes\left(\xi_2\otimes\eta_2\right)
\otimes\dots\,,
\label{sigma02}
\end{equation}
where $\xi$ ($\eta$) is the initial state of the subancilla having the same dimension as $S$ ($M$). In full analogy with \eq(\ref{sigma0-bi}), $\rho_0$ ($\bar\eta_M$) is the initial state of $S$ ($M$).

\subsection{System-ancilla collisions}

A further distinctive feature of the generalized CM with memory is that the collisions with the ancillas now involve $S$ as well. By definition, the collision between $S$-$M$ and the $n$th bipartite ancilla is described by the non-unitary {\it four}-partite CPT map
\begin{eqnarray}
  \mathcal{S}_{n} [\sigma] & = & p_n  \sigma + (1-p_n)\,\hat{V}_{Sn_1} \hat{S}_{Mn_2}
  \sigma \hat{S}_{Mn_2}^\dag \hat{V}_{Sn_1}^\dag\label{mapS}\,,
\end{eqnarray}
where $n_1$ and $n_2$ are the two $n$'s subancillas that are isodimensional to $S$ and $M$, respectively, while $\hat{V}_{Sn_1}$ is a unitary operator acting on $S$ and subancilla $n_1$. 
Map (\ref{mapS}) therefore swaps the states of $M$ and $n_2$ and, at the same time, applies the unitary $\hat{V}_{Sn_1}$ on $S$ and $n_1$, or leaves unchanged the state of $S$, $M$, $n_1$ and $n_2$ with probability $p_n$.
Note that, unlike the CM of the previous section [\cf\eq(\ref{PS})],
now we allow the probability $p_n$ to be in general step-dependent. The reason for this will become clear later on.

Based on \eq(\ref{mapS}) and the ancilla's initial state [\cf\eq(\ref{sigma02})], it is convenient to define a bipartite CPT map on $S$ and $M$ as
\begin{eqnarray}
  \widetilde{\mathcal{Z}}\,[ \rho_{{}_{\!S\!M}}]  & = & \tmop{Tr}_{n_1 n_2} \left\{
  \hat{V}_{Sn_1} \hat{S}_{Mn_2}\left( \rho_{{}_{\!S\!M}} \otimes \xi_{n_1} \otimes \eta_{n_2}\right)\,
  \hat{S}_{Mn_2}^\dag   \hat{V}_{Sn_1}^\dag \right\} \nonumber\\
  & = &\mathcal{Z}  [\tmop{Tr}_{M} \{\rho_{{}_{\!S\!M}}\}] \, \otimes\, \eta_{M}\,,\label{Ztilde}
\end{eqnarray}
where $\mathcal Z$ is the CPT map on $S$ defined by
\begin{eqnarray}
  \mathcal{Z} [ \rho] & = & \tmop{Tr}_{n_1} \left\{ \hat{V}_{Sn_1} \rho
  \otimes \xi_{n_1} \hat{V}_{Sn_1}^\dag \right\}\!. \label{Z}
\end{eqnarray}
The proof of the last step in \eq(\ref{Ztilde}) is given in Appendix A.

\eqs(\ref{Ztilde}) and (\ref{Z}) entail that the collision with the $n$th ancilla [see \eq(\ref{mapS})] changes the {\it reduced} state of $S$ and $M$, $\rho_{{}_{\!S\!M}}$, according to
\begin{eqnarray}
 {\rm Tr}_{n_1n_2} \left\{\mathcal{S}_{n} (\rho_{{}_{\!S\!M}}\,\xi_{n_1}\,\eta_{n_2})\right\} & = & p_n\, \rho_{{}_{\!S\!M}} + (1-p_n) \, \widetilde{\mathcal{Z}} [\rho_{{}_{\!S\!M}}]\\
 &=&p_n\,\rho_{{}_{\!S\!M}} + (1-p_n)  \,  \mathcal{Z} [ \tmop{Tr}_{M} \{\rho_{{}_{\!S\!M}}\}] \, \otimes\eta_M.\nonumber\,\,\,\,\,\,\,\,\,\,\,\,\,\,\label{mapS-red}
\end{eqnarray}
The essential effect of the collision, thereby, is to either leave with probability $p_n$ the $S$-$M$ state unchanged or, 
with probability $1{-}p_n$, to apply the CPT map $\mathcal Z$ on $S$ by simultaneously resetting the $M$'s state to $\eta$.

\subsection{Discrete dynamics}

Similarly to the CM in \rref\cite{lorenzoPRA2016} (see previous section), the initial state (\ref{sigma02}) evolves through an underlying $S$-$M$ unitary dynamics that is interrupted at each {\it fixed} time step $\tau$ by a collision described by \eq(\ref{mapS}) involving a fresh bipartite ancilla that is still in state $\xi \otimes \eta$. Accordingly, the overall state at the $n$th step is given by $\sigma_n=\mathcal S_n\,\mathcal \,U_\tau\,\mathcal S_{n-1}\,\mathcal U_\tau\,\ldots\mathcal S_2\,\mathcal U_\tau\,\mathcal S_1\,\mathcal U_\tau [\sigma_0]$.

Starting from $\rho_{S\!M}^{(0)}{=}\rho_0 \otimes \bar\eta_M$ [see \eq(\ref{sigma02})], at the end of the first step the reduced $S$-$M$ state is turned into
\begin{equation}
\rho^{(1)}_{S\!M}=\U\left[\rho^{(0)}_{S\!M}\right]\label{step12}\,.
\end{equation} 
Next, the collision with ancilla 1 described by map $\mathcal S_{1}$ [see \eq(\ref{mapS})] takes place followed by another application of the $S{-}M$ unitary. At the end of the second step, the $S$-$M$ state thus reads
\begin{eqnarray}
\rho^{(2)}_{S\!M}\!\!\!\!&{=}&\!\!\!\!{\rm Tr}_{1_11_2}\left\{\mathcal U_\tau\mathcal S_1\left[\rho^{(1)}_{SM}\,\xi_{1_1}\!\eta_{1_2}\right]\right\}{=}{\rm Tr}_{1_11_2}\left\{\left(p_1\mathcal U_\tau{+}q_1\mathcal U_\tau\tilde{\mathcal Z}\right)\left[\rho^{(1)}_{S\!M}\,\xi_{1_1}\!\eta_{1_2}\right]\right\}\nonumber\\
\!\!\!\!&{=}& \!\!\!\!p_1~\U\left[\rho^{(1)}_{S\!M}\right]{+} q_1 ~\U\tilde{\mathcal Z} \left[\rho^{(1)}_{S\!M}\right]\label{step22-a}\,,
\end{eqnarray}
where the trace is taken over the $n$th ancilla for $n=1$ and to simplify the notation we set $q_n=1{-}p_n$. By replacing in the last identity the state at the end of the first step (\ref{step12}), \eq(\ref{step22-a}) can be expressed as a function of the initial $S$-$M$ state only as
\begin{eqnarray}
  \rho^{(2)}_{S \hspace{-0.17em} M} & = & \underbrace{p_{1}
  \mathcal{U}^{2}_{\tau}}_{0 \, \tmop{jumps}} \left[ \rho^{(0)}_{S
  \hspace{-0.17em} M} \right] \hspace{-0.17em} + \hspace{-0.17em}
  \underbrace{q_{1} \mathcal{U}_{\tau}  \widetilde{\mathcal{Z}} 
  \mathcal{U}_{\tau}}_{1 \, \tmop{jump}} \left[ \rho^{(0)}_{S \hspace{-0.17em}
  M} \right] . \hspace{0.17em} \hspace{0.17em}  \label{step22}
\end{eqnarray}
Since the elapsed time of the process is an integer multiple of the time step
$\tau$ and given that a jump (if any) occurs at the end of each time step
$\tau$, at the second step either 0 or 1 jumps have taken place. The former
and latter cases correspond to the terms featuring zero or one
$\widetilde{\mathcal{Z}}$ in {\eq}(\ref{step22}) as highlighted by the
captions.
At the end of the 3rd step, after the application of maps $\mathcal{S}_{2}$
and $\mathcal{U}_{\tau}$, an analogous calculation leads to
\begin{eqnarray}
  \rho^{(3)}_{S \hspace{-0.17em} M} & = & p_{2} ~ \mathcal{U}_{\tau}  \left[
  \rho^{(2)}_{S \hspace{-0.17em} M} \right] +q_{2} ~ \mathcal{U}_{\tau} 
  \widetilde{\mathcal{Z}} \left[ \rho^{(2)}_{S \hspace{-0.17em} M} \right] \\
  & = & \underbrace{p_{2} p_{1} \mathcal{U}^{3}_{\tau}}_{0 \, \tmop{jumps}}
  \left[ \rho^{(0)}_{S \hspace{-0.17em} M} \right] \hspace{-0.17em} +
  \hspace{-0.17em} \underbrace{( p_{2} q_{1} \mathcal{U}^{2}_{\tau} 
  \widetilde{\mathcal{Z}}  \mathcal{U}_{\tau} +q_{2} p_{1} \mathcal{U}_{\tau} 
  \widetilde{\mathcal{Z}}  \mathcal{U}_{\tau}^{2} )}_{1 \, \tmop{jump}} \left[
  \rho^{(0)}_{S \hspace{-0.17em} M} \right]  \nonumber\\
  &  & + \underbrace{q_{2} q_{1}
  \mathcal{U}_{\tau}  \widetilde{\mathcal{Z}} \mathcal{U}_{\tau} 
  \widetilde{\mathcal{Z}}  \mathcal{U}_{\tau}}_{2 \, \tmop{jumps}} \left[
  \rho^{(0)}_{S \hspace{-0.17em} M} \right] \hspace{0.17em} , \hspace{0.17em}
  \hspace{0.17em} \hspace{0.17em} \hspace{0.17em} \hspace{0.17em}
  \hspace{0.17em} \hspace{0.17em} \hspace{0.17em} \hspace{0.17em}
  \hspace{0.17em} \label{step32} \nonumber
\end{eqnarray}
showing that, as expected, 0, 1 or 2 jumps are possible in this case
corresponding to as many applications of the map $\widetilde{\mathcal{Z}}$.
In a similar fashion, at the 4th step we get
\begin{eqnarray}
  \rho^{(4)}_{S \hspace{-0.17em} M} & = & p_{3} ~ \mathcal{U}_{\tau}  \left[
  \rho^{(3)}_{S \hspace{-0.17em} M} \right] +q_{3} ~ \mathcal{U}_{\tau} 
  \widetilde{\mathcal{Z}} \left[ \rho^{(3)}_{S \hspace{-0.17em} M} \right] \\
  & = & \underbrace{p_{3} p_{2} p_{1} \mathcal{U}^{4}_{\tau}}_{0 \,
  \tmop{jumps}} \left[ \rho^{(0)}_{S \hspace{-0.17em} M} \right]
  \hspace{-0.17em}  \nonumber\\
  &  & + \hspace{-0.17em} \underbrace{( p_{3} p_{2} q_{1}
  \mathcal{U}^{3}_{\tau}  \widetilde{\mathcal{Z}}  \mathcal{U}_{\tau}
       \piu p_{3}
  q_{2} p_{1} \mathcal{U}^{2}_{\tau}  \widetilde{\mathcal{Z}} 
  \mathcal{U}_{\tau}^{2} \piu q_{3} q_{2} p_{1} \mathcal{U}_{\tau} 
  \widetilde{\mathcal{Z}}  \mathcal{U}^{3}_{\tau} )}_{1 \, \tmop{jump}} \left[
  \rho^{(0)}_{S \hspace{-0.17em} M} \right] \nonumber\\
  &  & + \underbrace{(p_{3} q_{2} q_{1} \mathcal{U}^{2}_{\tau} 
  \widetilde{\mathcal{Z}} \mathcal{U}_{\tau}  \widetilde{\mathcal{Z}} 
  \mathcal{U}_{\tau} \piu q_{3} p_{2} q_{1} \mathcal{U}_{\tau} 
  \widetilde{\mathcal{Z}} \mathcal{U}^{2}_{\tau}  \widetilde{\mathcal{Z}} 
  \mathcal{U}_{\tau} \piu q_{3} q_{2} p_{1} \mathcal{U}_{\tau} 
  \widetilde{\mathcal{Z}} \mathcal{U}_{\tau}  \widetilde{\mathcal{Z}} 
  \mathcal{U}^{2}_{\tau})}_{2 \, \tmop{jumps}} \left[ \rho^{(0)}_{S
  \hspace{-0.17em} M} \right] \nonumber\\
  &  & + \underbrace{q_{3} q_{2} q_{1} \mathcal{U}_{\tau} 
  \widetilde{\mathcal{Z}} \mathcal{U}_{\tau}  \widetilde{\mathcal{Z}}
  \mathcal{U}_{\tau}  \widetilde{\mathcal{Z}}  \mathcal{U}_{\tau}}_{3 \,
  \tmop{jumps}} \left[ \rho^{(0)}_{S \hspace{-0.17em} M} \right]
  \hspace{0.17em} . \nonumber
\end{eqnarray}
In order to write down the $n$th-step state in a compact form, having
in mind the structure of \eq(\ref{dysonNM}), we first note that based
on \eq(\ref{mapU}) any $k$th power of the map $\mathcal U_\tau$ is given by $\mathcal U_\tau^k=\mathcal U_{k\tau}$ (in the following we will further set $\mathcal U_{k}\equiv\mathcal U_{k\tau}$ to simplify the notation).

By induction, the $n$th-step state for arbitrary $n{\ge}2$ is given by
\begin{eqnarray}
\rho^{(n)}_{S\!M}&=&\left(\prod_{\ell=1}^{n{-}1}p_\ell \right)\,\mathcal U_{n}\left[\rho^{(0)}_{S\!M}\right]+
\sum_{j=1}^{n-1} \sum_{k_j=1}^{n-1}\sum _{k_{j{-}1}=1}^{k_j-1} \ldots \sum _{k_1=1}^{k_2-1}\,\pi(k_j,\ldots,k_1)\nonumber\\&&
   \mathcal U_{n-k_j} \,\tilde{\mathcal Z}\, \mathcal U_{k_j{-}k_{j{-}1}} \tilde{\mathcal Z}\ldots \, \tilde{\mathcal Z}\,\mathcal U_{k_2-k_1}\,\tilde{\mathcal Z}\, \mathcal U_{k_1}\left[\rho^{(0)}_{S\!M}\right],\nonumber\\
   \label{formula}
\end{eqnarray}
where $\pi(k_j,\ldots,k_1)$ stands for the probability to perform exactly $j$ jumps at specific steps $\{k_j,\ldots,k_1\}$ and reads
\begin{equation}
\pi(k_j,\ldots,k_1){=}\left(\prod_{\ell{=}k_j\piu1}^{n\meno1}\!\!p_\ell \right)\!\!q_{k_j}\!\!\left(\prod_{\ell{=}k_{j\meno1}\piu1}^{k_{j}\meno1}\!\!\!\!p_\ell \right)\!\!q_{k_{j\meno1}}\!\ldots \!q_{k_{2}}\!\!\left(\prod_{\ell{=}k_1\piu1}^{k_{2}\meno1}p_\ell \right)\!q_{k_{1}}\!\!\left(\prod_{\ell{=}1}^{k_{1}{-}1}p_\ell \right).
\label{jumps-prob}\end{equation}

\subsection{Continuous-time limit}

In order to perform the continuous-time limit, in analogy with
{\eq}(\ref{p-Gamma}) we introduce the quantities
\begin{eqnarray}
  p ( t_{k} -t_{k-1} ) = \mathe^{- \int^{t_{k} -t_{k-1}}_{0} \mathd s \phi ( s
  )} , &  & q ( t_{k} -t_{k-1} ) =1- \mathe^{- \int^{t_{k} -t_{k-1}}_{0}
  \mathd s \phi ( s )} , \nonumber
\end{eqnarray}
corresponding respectively to the probability of no jump or one jump to take
place in each small time interval $t_{k} -t_{k-1}$, which in the case of
constant hazard function $\phi ( s )$ reduces to a Poisson distribution for the jumps. According to the definition of a renewal process, the jump
probabilities thereby depend only on the elapsed time.
In this representation, the various contributions appearing in
\eq(\ref{jumps-prob}), in the limit of a large number of short steps such that the
time intervals between steps become increasingly small, can be written as
\begin{eqnarray}
  \left( \prod^{j}_{\ell =k_{} +1} 
  p_{\ell} \right)  \,
  \, q_{k} & = & \prod^{j}_{\ell =k_{} +1} \,
  \, \, \, \,
  \mathe^{- \int^{t_{l} -t_{l-1}}_{0} \mathd s \phi ( s )} \left( 1- \mathe^{-
  \int^{t_{k} -t_{k-1}}_{0} \mathd s \phi ( s )} \right) \nonumber\\
  & \approx & \mathe^{- \int^{t_{j} -t_{k}}_{0} \mathd s \phi ( s )} -
  \mathe^{- \int^{t_{j} -t_{k_{} -1}}_{0} \mathd s \phi ( s )} \nonumber\\
  & \approx & \phi ( t_{j} -t_{k-1} ) \mathe^{- \int^{t_{j} -t_{k-1}}_{0}
  \mathd s \phi ( s )} ( t_{k} -t_{k-1} )\,.\nonumber
\end{eqnarray}
This shows that the function $\phi ( t )$ has indeed the role of hazard rate function [\cf
\eq(\ref{f-and-g}], which determines the renewal process describing the time distribution
of jumps. Hence, we can thus finally identify
\begin{eqnarray}
  \left( \prod^{j}_{\ell =k_{} +1} \, \,
  \, \, p_{\ell} \right)  \,
  \, q_{k} \approx f (t_{j} -t_{k-1} ) \tmop{dt}_{k-1}\, . 
\end{eqnarray}
The first term in (\ref{formula}) accordingly becomes
\begin{eqnarray}
  \prod_{\ell =1}^{k} p_{l} \rightarrow e^{- \int_{0}^{t_{k}} \phi (s) ds}
  =g (t_{k} )\,\, . 
\end{eqnarray}
Therefore, \eq (\ref{formula}) in the continuous-time limit reads
\begin{eqnarray}
  \rho_{S M} (t) & = & g (t)  \mathcal{U}_{t} [ \rho_{S
  M} (0)] 
\nonumber
 \\
&  & 
+ \sum_{j=1}^{\infty} \int_{0}^{t} \,
  \, \tmop{dt}_{j}   \ldots \, \,
  \int_{0}^{t_{2}} \, \, \tmop{dt}_{1} \hspace{0.27em} f (t-t_{j} )
                          \ldots f (t_{2} -t_{1} ) g (t_{1} ) 
\nonumber
 \\
&  & \hphantom{ \mathcal{U}_{t} [ \rho_{SM} (0)] }  
  \times \mathcal{U}_{t-t_{j}}
  \hspace{0.17em} \widetilde{\mathcal{Z}}  \hspace{0.17em} \ldots
  \hspace{0.17em} \widetilde{\mathcal{Z}}  \hspace{0.17em} \mathcal{U}_{t_{2}
  -t_{1}}  \hspace{0.17em} \widetilde{\mathcal{Z}}  \hspace{0.17em}
  \mathcal{U}_{t_{1}} [ \rho_{SM} (0)]\,\, .  
\label{formula2}
\end{eqnarray}

\subsection{Reduced dynamics}

So far, we have focused on the bipartite system $S$-$M$, working out its
evolution. We now consider the resulting reduced dynamics for the system $S$,
which embodies the degrees of freedom of the open quantum system of interest.
We first recall that $\rho_{S \hspace{-0.17em} M} (0) = \rho_{0} \otimes
\bar{\eta}_{M}$ [see {\eq}(\ref{sigma02})], which ensures the existence of the
reduced dynamical map of $S$. When this expression is replaced in \eq
(\ref{formula2}) upon taking the trace over $M$ we get
\begin{eqnarray}
  \rho (t) & = & g (t) \tmop{Tr}_{M} \{ \mathcal{U}_{t} [ \rho_{0} \otimes
  \bar{\eta}_{M} ] \} \nonumber\\
  &  & + \sum_{j=1}^{\infty} \int_{0}^{t} \hspace{-0.17em} \hspace{-0.17em}
  \tmop{dt}_{j}   \ldots \hspace{-0.17em} \hspace{-0.17em} \int_{0}^{t_{2}}
  \hspace{-0.17em} \hspace{-0.17em} \tmop{dt}_{1}   \hspace{0.27em} f (t-t_{j}
  ) \ldots f (t_{2} -t_{1} ) g (t_{1} ) \nonumber\\
  &  & \hspace{0.17em} \hspace{0.17em}
  \hspace{0.17em} \hspace{0.17em} 
  \hspace{0.17em} \hspace{0.17em} \hspace{0.17em} \hspace{0.17em}
  \hspace{0.17em} \hspace{0.17em} \hspace{0.17em} \hspace{0.17em}
  \hspace{0.17em} \hspace{0.17em} \hspace{0.17em} \hspace{0.17em}
  \hspace{0.17em} \hspace{0.17em} \hspace{0.17em} \hspace{0.17em}
  \hspace{0.17em} \hspace{0.17em} \hspace{0.17em} \times \tmop{Tr}_{M} \left\{
  \mathcal{U}_{t-t_{j}}  \hspace{0.17em} \widetilde{\mathcal{Z}} 
  \hspace{0.17em} \ldots \hspace{0.17em} \widetilde{\mathcal{Z}} 
  \hspace{0.17em} \mathcal{U}_{t_{2} -t_{1}}  \hspace{0.17em}
  \widetilde{\mathcal{Z}}  \hspace{0.17em} \mathcal{U}_{t_{1}} [ \rho_{0}
  \otimes \bar{\eta}_{M} ] \right\} \hspace{0.17em}\!.\,\,\, 
\end{eqnarray}
By next introducing, according to {\eq}(\ref{special-case}), the CPT maps
$\mathcal{E}_{t}$ and $\bar {\mathcal{E}}_{t}$, whose definition is thus
identical to the model in {\rref}{\cite{lorenzoPRA2016}}, and recalling
{\eqs}(\ref{Ztilde}) and (\ref{Z}), we get
\begin{eqnarray}
  \rho (t) & = & g (t)  \bar {\mathcal{E}}_{t} [ \rho_{0} ] +
  \sum_{j=1}^{\infty} \int_{0}^{t} \hspace{-0.17em} \hspace{-0.17em}
  \tmop{dt}_{j}   \ldots \hspace{-0.17em} \hspace{-0.17em} \int_{0}^{t_{2}}
  \hspace{-0.17em} \hspace{-0.17em} \tmop{dt}_{1}   \hspace{0.27em} f (t-t_{j}
  ) \ldots f (t_{2} -t_{1} ) g (t_{1} ) \nonumber\\
  &  & \hphantom{ \bar {\mathcal{E}}_{t} [ \rho_{0} ]}\times \tmop{Tr}_{M} \left\{
  \mathcal{U}_{t-t_{j}}  \hspace{0.17em} \widetilde{\mathcal{Z}} 
  \hspace{0.17em} \ldots \hspace{0.17em} \widetilde{\mathcal{Z}} 
  \hspace{0.17em} \mathcal{U}_{t_{2} -t_{1}}  [ \mathcal{Z} [
  \bar {\mathcal{E}}_{t} [ \rho_{0} ] ] \otimes \eta_{M} ] \right\} . \label{rhoz}
\end{eqnarray}
The argument of the partial trace can be expressed by iteration according to
\begin{multline}
  \mathcal{U}_{t-t_{j}}  \hspace{0.17em} \widetilde{\mathcal{Z}} 
  \hspace{0.17em} \ldots \hspace{0.17em} \widetilde{\mathcal{Z}} 
  \hspace{0.17em} \mathcal{U}_{t_{2} -t_{1}}  \left[ \mathcal{Z} \left[
  \bar {\mathcal{E}}_{t} \hspace{0.17em} [ \rho_{0} ] \right] \otimes
  \eta_{M} \right] \\ =\mathcal{U}_{t-t_{j}}  \hspace{0.17em}
  \widetilde{\mathcal{Z}}  \hspace{0.17em} \ldots \hspace{0.17em} \left[
  \mathcal{Z}  \left[ \hspace{0.17em} \mathcal{E}_{t_{2} -t_{1}}  \left[ \mathcal{Z} 
  \hspace{0.17em} \left[ \bar {\mathcal{E}}_{t} \hspace{0.17em} [ \rho_{0} ]
  \right]\right]\right] \otimes \eta_{M} \right] , \nonumber
\end{multline}
which finally leads to the expression
\begin{eqnarray}
  \mathrm{Tr}_{M} \left\{ \mathcal{U}_{t-t_{j}}  \hspace{0.17em}
  \widetilde{\mathcal{Z}}  \hspace{0.17em} \ldots \hspace{0.17em}
  \widetilde{\mathcal{Z}}  \hspace{0.17em} \mathcal{U}_{t_{2} -t_{1}}  \left[
  \mathcal{Z} \left[ \bar {\mathcal{E}}_{t} \hspace{0.17em} [ \rho_{0} ]
  \right] \otimes \eta_{M} \right] \right\} = \mathcal{E}_{t-t_{j}} 
  \hspace{0.17em} \mathcal{Z}  \hspace{0.17em} \ldots \hspace{0.17em}
  \mathcal{Z}  \hspace{0.17em} \mathcal{E}_{t_{2} -t_{1}}  \mathcal{Z} 
  \hspace{0.17em} \bar {\mathcal{E}}_{t} \hspace{0.17em} [ \rho_{0} ] ,
  \nonumber
\end{eqnarray}
where for the sake of simplicity we have removed the nested square
brackets in the last expression.  When this result is replaced in
{\eq}(\ref{rhoz}), we end up with {\eq}(\ref{dysonNM}). Accordingly,
the reduced dynamics of $S$ in the continuous-time limit necessarily
obeys ME (\ref{ME-piecewise}) with no restrictions.

We can therefore conclude that the generalized collision model with memory
constructed here is indeed able to reproduce altogether the piecewise NM
dynamics with jumps considered in {\rref}{\cite{vacchiniPRL2016}}.

\section{Conclusions and outlook}\label{conc}
\setcounter{equation}{0}

The Gorini-Kossakowski-Lindblad-Sudarshan ME has been for over 40 years
the workhorse of open quantum systems theory. It  
embodies the basic reference for open dynamics that lack
memory effects. Clearly, though, in the case of strong
coupling and/or structured reservoirs a memoryless Markovian description
fails to faithfully capture the relevant features of the
dynamics. Many non-trivial challenges follow, in particular the need for 
more general {\it evolution equations} that ensure a well-defined 
(i.e., CPT) dynamics and, at the same time, effectively describe memory effects. 
On top of this, it is highly desirable that these theoretical descriptions be 
associated with corresponding environmental models 
thus providing an underlying {\it microscopic} interpretation and, possibly, a 
controlled implementation of such non-Markovian
dynamics.  

Both the above aspects were the focus of this paper. Starting from
a recently proposed family of memory-kernel MEs corresponding to a large 
class of generally non-Markovian time evolutions, we showed that {\it any} such ME
admits a microscopic CM from which it can be obtained as the equation governing 
its continuous-time-limit reduced dynamics. 

Specifically, the considered time
evolutions consist of piecewise dynamics in which a
continuous, generally non-Markovian, time evolution is interrupted at
random times, distributed according to a general waiting time
distribution, by a quantum jump described by a general CPT
transformation. These dynamics obey a closed memory-kernel ME. 
In this work, we showed that one such ME can be obtained as the continuous-time limit of a CM 
where memory effects are due to auxiliary degrees
of freedom (which we indeed called memory) mediating the action of the
environment on the system. As a distinctive feature of the CM, each bath ancilla is 
{\it bipartite} comprising a pair of subancillas. Each collision occurs in the form of a map that with some probability
swaps the state of the memory and one subancilla, while a unitary is at the same time applied 
on the system under study and the other subancilla. 
As a further hallmark of the considered CM, the probability for such swap-and-unitary operation can  
{\it depend} at will on the time step.

As remarked in the main text, the ancillas' doubling along with the step-dependance of the aforementioned probability
are the crucial features marking the difference between the CM in \rref\cite{lorenzoPRA2016} and the one addressed here (which can thus be viewed as a non-trivial generalization of the former). They allow to introduce
a jump map as well as a waiting time distribution of arbitrary shape.

It is interesting to note that the term ``collisional model" was at times used in the literature (see \eg \cite{Budini2004a}) to denote a quantum dynamics that is interrupted at random times by ``collisions" -- that is jumps in fact -- just like in the framework addressed in \rref\cite{vacchiniPRL2016}. In this respect, our work provides a connection between this definition of CM, based on {\it random}-time collisions, and the one used throughout the paper, where instead collisions occur at  {\it fixed} times.

We finally point out that ME (\ref{ME-piecewise}) was obtained in \rref\cite{vacchiniPRL2016} within a general framework based on the quantization of a family of classical stochastic dynamics. Since this quantization involves {\it non-commuting} operators, ME (\ref{ME-piecewise}) arises only as one of {\it two} possible cases corresponding to different operator orderings. The question whether or not a class of underlying CMs can be devised even for the ME arising in the other case \cite{vacchiniPRL2016} -- which is qualitatively different from ME (\ref{ME-piecewise}) --  is under ongoing investigations.

\section*{Appendix A}
\def\theequation{A.\arabic{equation}}
\setcounter{equation}{0}
We here provide the proof of the last identity in {\eq}(\ref{Ztilde}). Let us
first recall the starting point, namely the definition of the map
$\widetilde{\mathcal{Z}}$ given in the first line of
{\eq}(\ref{Ztilde}), omitting the tensor product symbol to simplify
the notation
\begin{eqnarray}
  \widetilde{\mathcal{Z}} [ \rho_{SM} ] & =
  & \tmop{Tr}_{n_{1} n_{2}} \left\{ \hat{V}_{Sn_{1}}  \hat{S}_{Mn_{2}} 
  \rho_{SM}  \xi_{n_{1}} 
  \eta_{n_{2}} \hspace{0.17em} \hat{V}_{Sn_{1}}^{\dag} 
  \hat{S}_{Mn_{2}}^{\dag} \right\} \hspace{0.17em} ,  \label{Ztilde2}
\end{eqnarray}
and consider two orthonormal bases $\{| \mu \rangle_{M} \}$ and $\{| \nu
\rangle_{n_{2}} \}$ in the Hilbert spaces of $M$ and $n_{2}$, respectively. In
terms of these vectors, the swap operator $\hat{S}_{Mn_{2}}$ is expressed as
\[ \hat{S}_{Mn_{2}} = \sum_{\mu , \nu} | \mu \rangle \langle \nu |_{M} \otimes
   | \nu \rangle \langle \mu | \nobracket_{n_{2}} \hspace{0.17em} . \]
Using this expression in {\eq}(\ref{Ztilde2}) the rhs explicitly reads
\begin{eqnarray}
&& 
\hspace{-.6truecm}
\sum\limits_{\substack{\mu , \nu\\ \mu' , \nu'}}
\tmop{Tr}_{n_{1} n_{2}} \{ \hat{V}_{Sn_{1}} \nobracket | \mu
  \rangle \langle \nu |_{M} \otimes | \nu \rangle \langle \mu |
  \nobracket_{n_{2}} \rho_{SM} 
  \xi_{n_{1}}  \eta_{n_{2}} | \nu' \rangle \langle \mu' |_{M} \otimes
  | \mu' \rangle_{} \langle \nobracket \nu' |_{n_{2}} \nobracket
  \hat{V}_{Sn_{1}}^{\dag} \}= \nonumber\\
  & &
\hspace{-.6truecm}
\sum\limits_{\substack{\mu , \nu\\ \mu' , \nu'}}
 \tmop{Tr}_{n_{1} n_{2}} \{ \hat{V}_{Sn_{1}} | \mu \rangle_{M}
  \otimes | \nu \rangle_{n_{2}} \hspace{-0.17em} \langle \nu | \nobracket
  \rho_{SM} | \nu' \rangle_{M} 
  \xi_{n_{1}}  \hspace{-0.17em} \langle \mu | \nobracket \eta_{n_{2}} |
  \mu' \rangle_{n_{2}} \nobracket \hspace{0.17em}_{M} \hspace{-0.17em} \langle
  \mu' | \hspace{0.17em}_{n_{2}} \hspace{-0.17em} \langle \nobracket \nu' |
  \hat{V}_{Sn_{1}}^{\dag} \} , \nonumber
\end{eqnarray}
so that taking the partial trace over $n_{2}$ we end up with
\begin{displaymath}
  \widetilde{\mathcal{Z}} [ \rho_{\hspace{-0.17em} S \hspace{-0.17em}
    M} ]
=
\sum_{\mu , \mu'} \tmop{Tr}_{n_{1}} \{ \hat{V}_{Sn_{1}} | \mu \rangle_{M}
  \hspace{0.17em} \hspace{0.17em} \hspace{-0.17em} \tmop{Tr}_{M}
  \{\rho_{SM} \}  \xi_{n_{1}} \langle \mu | \nobracket
  \eta_{n_{2}} | \mu' \rangle_{n_{2}} \nobracket \hspace{0.17em}_{M}
  \hspace{-0.17em} \langle \mu' | \hspace{0.17em} \hat{V}_{Sn_{1}}^{\dag}
  \} . \nonumber
\end{displaymath}
By recalling that the state $\rho$ of the reduced system is just the marginal
of $\rho_{SM}$ and by noting that the
expression in square brackets swaps $\eta_{n_{2}}$ and $\eta_{M}$, we
finally get
\begin{eqnarray*}
  \widetilde{\mathcal{Z}} [ \rho_{SM} ] & =& \tmop{Tr}_{n_{1}} \{ \hat{V}_{Sn_{1}} \rho \xi_{n_{1}} 
  \eta_{M}  \hat{V}_{Sn_{1}}^{\dag} \} 
\\
&=& \tmop{Tr}_{n_{1}} \left\{
  \hat{V}_{Sn_{1}} \rho \hspace{0.17em} \xi_{n_{1}} 
  \hat{V}_{Sn_{1}}^{\dag} \right\} \otimes \eta_{M} = \mathcal{Z} [ \rho ]
  \otimes \eta_{M} \hspace{0.17em} , \nonumber
\end{eqnarray*}
which according to the definition (\ref{Z}) of the map $\mathcal{Z}$ concludes
the proof.

\section*{Acknowledgments}
We acknowledge support from the EU Project QuPRoCs (Grant Agreement 641277) and the Fulbright Commission.

\end{document}